\documentclass[
    ,final            % use final for the camera ready runs
  ]
  {aipproc}

\layoutstyle{6x9}

\def\Journal#1#2#3#4{{#1} {\bf #2}, #3 (#4)}

\def\NIMA{{\em Nucl. Instrum. Methods} A}
\def\NPB{{\em Nucl. Phys.} B}
\def\PLB{{\em Phys. Lett.}  B}

\def\ZPC{{\em Z. Phys.} C}
\def\etal{{\sl et al.}}
\begin{document}

\title{NuTeV Structure Function Measurement }



\author{M.~Tzanov for the NuTeV Collaboration%$^7$, T. Adams$^4$, A.~Alton$^4$,
%  S.~Avvakumov$^8$, L.~de~Barbaro$^5$, P.~de~Barbaro$^8$,
%  R.~H.~Bernstein$^3$, A.~Bodek$^8$, T.~Bolton$^4$, S.~Boyd$^7$, 
%  J.~Brau$^6$, D.~Buchholz$^5$, H.~Budd$^8$, L.~Bugel$^3$, J.~Conrad$^1$,
%  R.~B.~Drucker$^6$, B.~T. Fleming$^1$, J.~Formaggio$^1$, R.~Frey$^6$,
%  J.~Goldman$^4$, M.~Goncharov$^4$, D.~A.~Harris$^8$, J.~H.~Kim$^1$,
%  S.~Koutsoliotas$^1$, R.~A.~Johnson$^2$, M.~J.~Lamm$^3$,
%  W.~Marsh$^3$, D.~Mason$^6$, J.~McDonald$^7$, K.~S.~McFarland$^8$, C.~McNulty$^1$,
%  D.~Naples$^7$, P.~Nienaber$^3$, V.~Radescu$^7$, A.~Romosan$^1$,
%  W.~K.~Sakumoto$^8$,H.~Schellman$^5$, M.~H.~Shaevitz$^1$,
%  P.~Spentzouris$^1$, E.~G.~Stern$^1$, N.~Suwonjandee$^2$,
%  N.~Tobien$^3$, A.~Vaitaitis$^1$, M.~Vakili$^2$,
%  U.~K.~Yang$^8$, J.~Yu$^3$, G.~P.~Zeller$^5$,
%  E.~D.~Zimmerman$^1$\\
%  {\it The NuTeV Collaboration}
}{
  address={\it %$^1$Columbia University, New York,NY, 
%               $^2$University of Cincinnati, Cincinnati, OH, 
%               $^3$Fermi National Accelerator Laboratory, Batavia, IL, 
%               $^4$Kansas State University, Manhattan, KS, 
%               $^5$Northwestern University, Evanston, IL,
%               $^6$University of Oregon, Eugene, OR, 
                University of Pittsburgh, Pittsburgh, PA 15260
%		$^8$University of Rochester, Rochester, NY.
 }
}

%\pub{Received (Day Month Year)}{Revised (Day Month Year)}

\begin{abstract}
The NuTeV experiment obtained high statistics samples of neutrino and
antineutrino charged current events during the 1996-1997 Fermilab fixed
target run. The experiment combines sign-selected neutrino and
antineutrino beams and the upgraded CCFR iron-scintillator neutrino
detector. A precision continuous calibration beam was used to determine
the muon and hadron energy scales to a precision of 0.7\% and 0.43\%
respectively. The structure functions $F_2(x,Q^2)$ and $xF_3(x,Q^2)$ 
obtained by fitting the y-dependence of the sum and the difference of 
the $\nu$  and $\overline \nu$ differential cross sections are presented.
\end{abstract}

\maketitle
%\section{Introduction} %) A SECTION HEADING
Neutrino deep inelastic scattering (DIS) provides a unique information
for the structure of the proton and QCD, allowing the measurement of two 
structure functions (SF): $F_2(x,Q^2)$, and the parity-violating 
$xF_3(x,Q^2)$,which is accessible only by neutrino DIS~\cite{nuRev}.
The NuTeV experiment is a high-energy fixed target $\nu-Fe$ 
scattering experiment, which combines two new features:
Separate high-purity neutrino and antineutrino beams,
used to tag the primary lepton in charged-current interactions, 
and a continuous precision calibration beam, which improves
the experiment's knowledge of the absolute energy scale
for hadrons and muon, produced in neutrino interactions,
to a precision of 0.43\% and 0.7\% respectively~\cite{nutCal}. 
NuTeV took data during 1996-97 and collected 
$8.6\times10^5$ $\nu$ and $2.4\times10^5$ $\overline{\nu}$ 
charged-current (CC) interactions that passed analysis cuts.

\section{$\nu$-Fe Charge Current Differential Cross Section}

The differential cross section is determined from
\begin{equation}
  \frac{d^2\sigma^{\nu(\overline{\nu})}}{dxdy} =  \frac{1}{\Phi(E)}\frac{d^2 N^{\nu(\overline{\nu})}(E)}{dxdy}, 
\label{eq:ccxsec}
\end{equation}
where $\Phi(E)$ is the $\nu (\overline{\nu})$ flux in energy bins. 
The cross section event sample is required to pass fiducial volume 
cuts, $\mu$ track reconstruction quality cuts,
a minimum muon energy
threshold $E_\mu > 15$ GeV, a minimum hadronic energy threshold $E_{HAD} > 10$~GeV,
and a minimum neutrino energy threshold $E_\nu > 30$~GeV.
%and minimum 
%energy thresholds of $E_\mu>15$~GeV, hadronic energy, 
%$E_{\scriptstyle HAD}>10$~GeV, and neutrino energy, $E_\nu>30$~GeV.
Selected events are binned in $x$, $y$, and $E_\nu$
bins, and corrected for acceptance and smearing using a fast
detector simulation. 
$Q^2>1$~GeV$^2$ is required 
to minimize the non-perturbative contribution to the cross section.
NuTeV data ranges from $10^{-3}$ to $0.95$ in $x$,
$0.05$ to $0.95$ in y, and from 30~GeV to 360~GeV in $E_\nu$.

The flux is determined from data with $E_{\scriptstyle
HAD}<20$ GeV using the ``fixed $\nu_0$'' relative flux 
extraction method~\cite{nuRev}.
The integrated number of events in this sample is proportional to the flux 
as $y=\frac{E_{\scriptscriptstyle HAD}}{E_\nu}\rightarrow 0$. Corrections up to order $y^2$, determined from the data sample,
%$\left(\frac{E_{\scriptscriptstyle HAD}}{E_\nu}\right)^2$
are applied to determine the relative flux to about the $1\%$ level. 
Flux is normalized using the world average $\nu$-Fe cross
section $\frac{\sigma^\nu}{E_\nu}=0.677 \times 10^{-38} cm^2/GeV$~\cite{pdg}.

The fast detector simulation, which takes into account acceptance and
resolution effects, uses an empirically determined set of PDFs
extracted by fitting the differential cross section~\cite{bg}.
The procedure is then iterated until convergence is achieved (within 3
iterations). 
Detector response functions are parameterized from the
NuTeV calibration beam data samples~\cite{nutCal}.

\section{Structure Functions}

\begin{figure}
\includegraphics[height=.35\textheight]{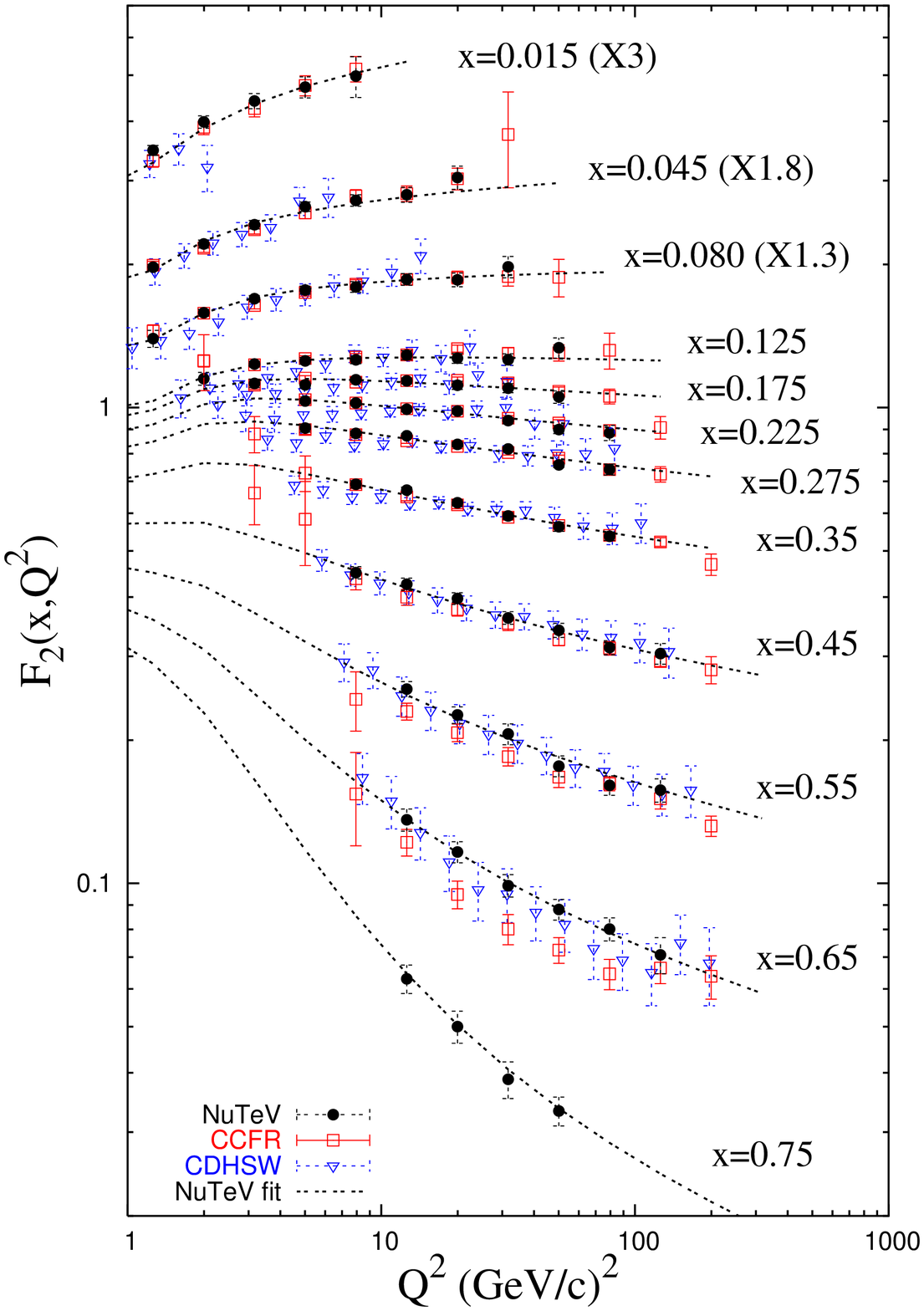}
\includegraphics[height=.35\textheight]{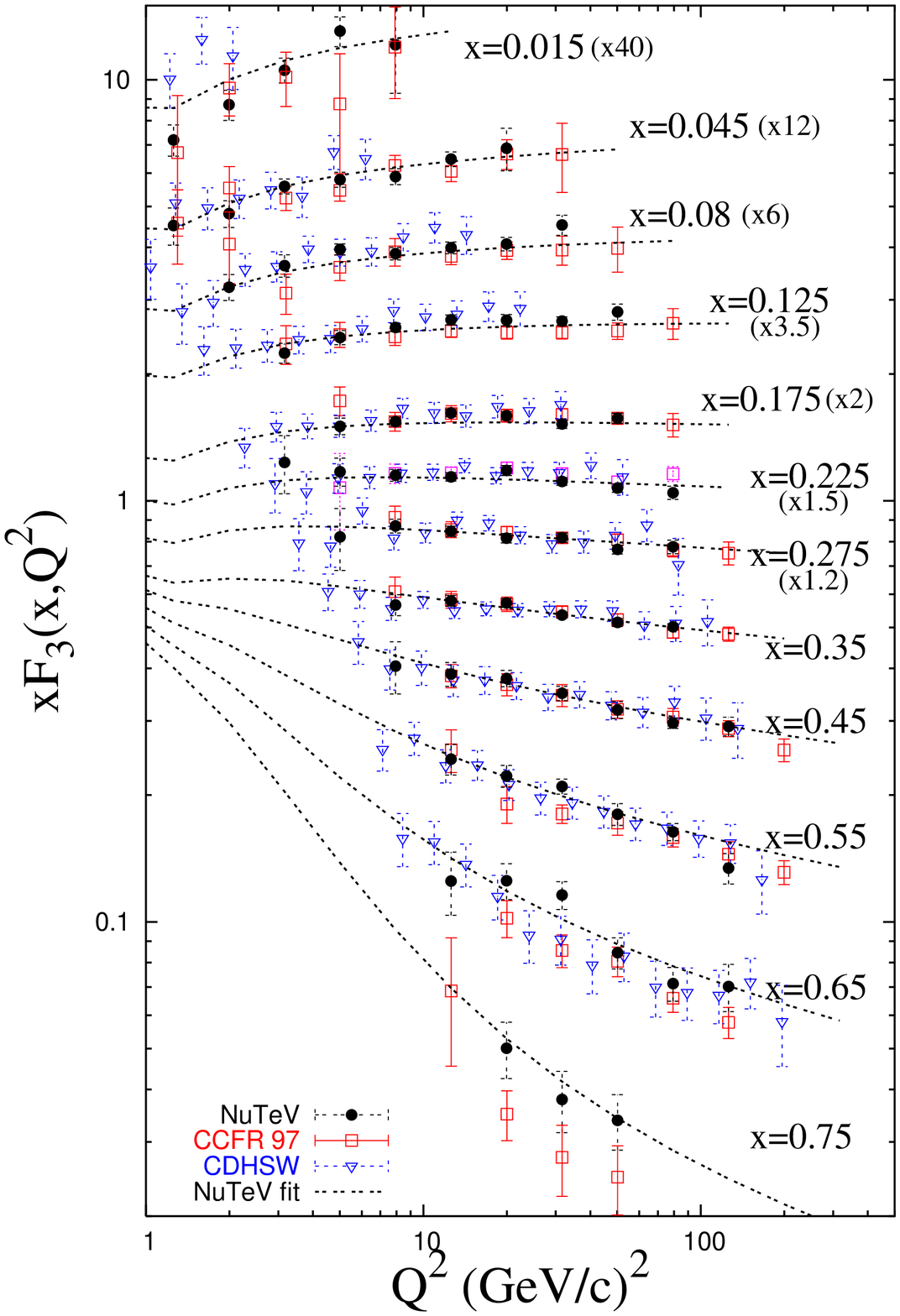}
  \caption{NuTeV $F_2$ (left) and $xF_3$ (right) in comparison with previous $\nu$-Fe scattering experiments.}
 \label{f2} 
\end{figure}

The structure function $F_2(x,Q^2)$ is determined from a fit to the y-dependence of the sum of the $\nu,\overline\nu$ differential cross sections:

\begin{equation}
\Big({\frac{d^2\sigma}{dx dy}}^{\nu}+{\frac{d^2\sigma}{dx dy}}^{\overline\nu}\Big)
%& =&
={\frac{G_{F}^2
M E}{\pi}}\Big[2\Big(1-y-{\frac{M x y}{2E}} 
+ {\frac{y^2}{2}{\frac{1+4 M^2 x^2/ Q^2}
         {1+R_{L}}}}\Big) F_{2} + {y} \Big(1 - {\frac{y}{2}}\Big)
         \Delta xF_{3}\Big],
\end{equation}
where $F_{2}=\frac{F_{2}^{\nu}+F_{2}^{\overline {\nu}}}{2}$,
$R_L(x,Q^2)$ is the ratio of the cross section for scattering
from longitudinally to transversely polarized W-bosons,% where
%$ 2xF_1= F_2\frac{1+{\frac{4M^2x^2}{Q^2}}}{(1 + R_L(x,Q^2))}$,
and $\Delta xF_3=xF_3^{\nu}-xF_3^{\overline{\nu}}$.
Cross sections are corrected for QED radiative effects and 
for 5.67\% excess of neutrons over protons in our iron target
before the sum is formed~\cite{bardin}.
%fits are performed.
To extract $F_2(x,Q^2)$ we use
$\Delta xF_3$ from a NLO QCD model as input (TRVFS)~\cite{trvfs}.
The input value of $R_L(x,Q^2)$ comes from
a fit to the world's measurements~\cite{rworld}.
NuTeV $F_2(x,Q^2)$ for neutrino scattering on iron is shown on Fig. \ref{f2} (left)
compared with previous $\nu$-Fe scattering measurements (CDHSW~\cite{cdhsw}, CCFR~\cite{ccfrxsec}).
NuTeV $F_2$ is in reasonable agreement with CDHSW and CCFR for
$x<0.4$. At high-$x$ NuTeV $F_2$ is systematically above CCFR: 4\% at $x=0.45$, 
9\% at $x=0.55$, 18\% at $x=0.65$.

Similarly, the structure function $xF_3(x,Q^2)$ is determined from a fit to the $y$-dependence  of the difference of the $\nu,\overline\nu$ differential cross sections:
\begin{equation}
\left[\frac{d^2\sigma^{\nu}}{dxdy} -
      \frac{d^2\sigma^{\overline{\nu}}}{dxdy}\right]=
\frac{2 G_F^2 M E}{\pi}
\left(y-\frac{y^2}{2}\right) xF_3^{\scriptstyle AVG}(x,Q^2),
\end{equation}
where $xF_3^{\scriptstyle AVG}=\frac{1}{2}(xF_3^{\nu}+xF_3^{\overline{\nu}})$.
$F^{\nu}_2(x,Q^2)\approx
F^{\overline{\nu}}_2(x,Q^2)$ are nearly identical so no
additional model input is required. 
Cross sections are corrected for QED radiative effects and 
for 5.67\% excess of neutrons over protons in our iron target
before the difference is formed~\cite{bardin}.
Fig. \ref{f2} (right) 
shows the NuTeV measurement of $xF_3(x,Q^2)$ compared to previous $\nu$-Fe 
results (CDHSW~\cite{cdhsw}, CCFR(97)~\cite{pdg}).
NuTeV $xF_3$ agrees with CCFR(97) and CDHSW for $x<0.4$. For $x>0.4$ NuTeV result
 is systematically higher than CCFR(97)~\cite{pdg}.
 
We have determined that the largest contribution to the discrepancy with CCFR at high-$x$
 is due to a mis-calibration of the magnetic field map
of the muon spectrometer in CCFR. NuTeV and CCFR used the same muon spectrometer.
Hence, the radial dependence of the magnetic field should be the same.
NuTeV mapped the entire surface of the muon spectrometer with calibration
beam of muons, which provided precise calibration of the magnetic field~\cite{nutCal}, while CCFR used a model for the magnetic field map and one high statistics calibration muon run, aimed at a single point of the spectrometer, to set the overall 
normalization~\cite{ccfrnim}.
The difference of the two magnetic field maps is an effective 0.8\% shift of
the muon energy scale, which accounts for a third of the discrepancy. Additional contributions to the discrepancy are the different cross section models used by
NuTeV and CCFR (3\% of the 18\%), and the NuTeV's improved muon and hadron energy 
smearing models (2\% of the 18\%). All of the above differences account 
for two thirds of the discrepancy.

 \begin{figure}
   \includegraphics[height=.36\textheight]{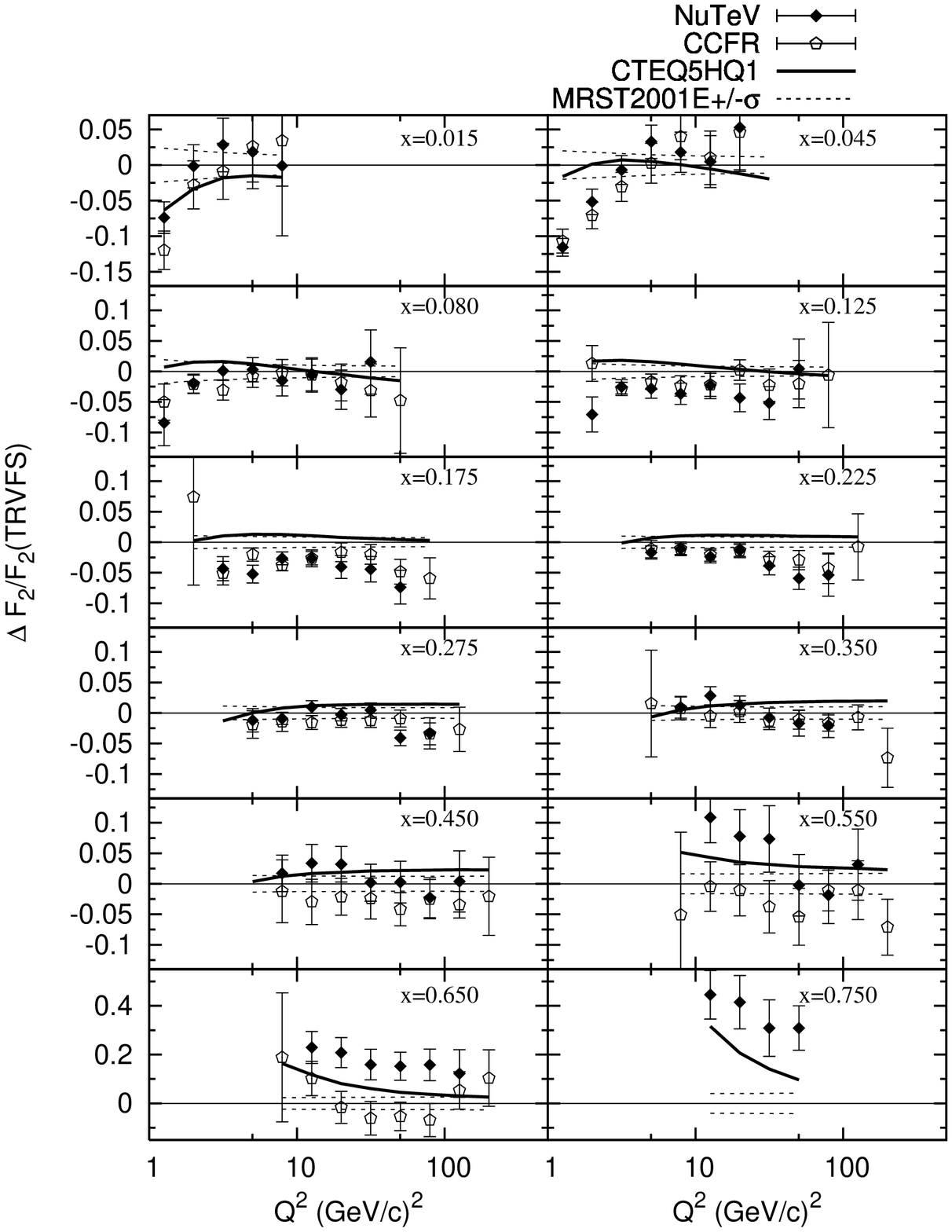}
   \includegraphics[height=.36\textheight]{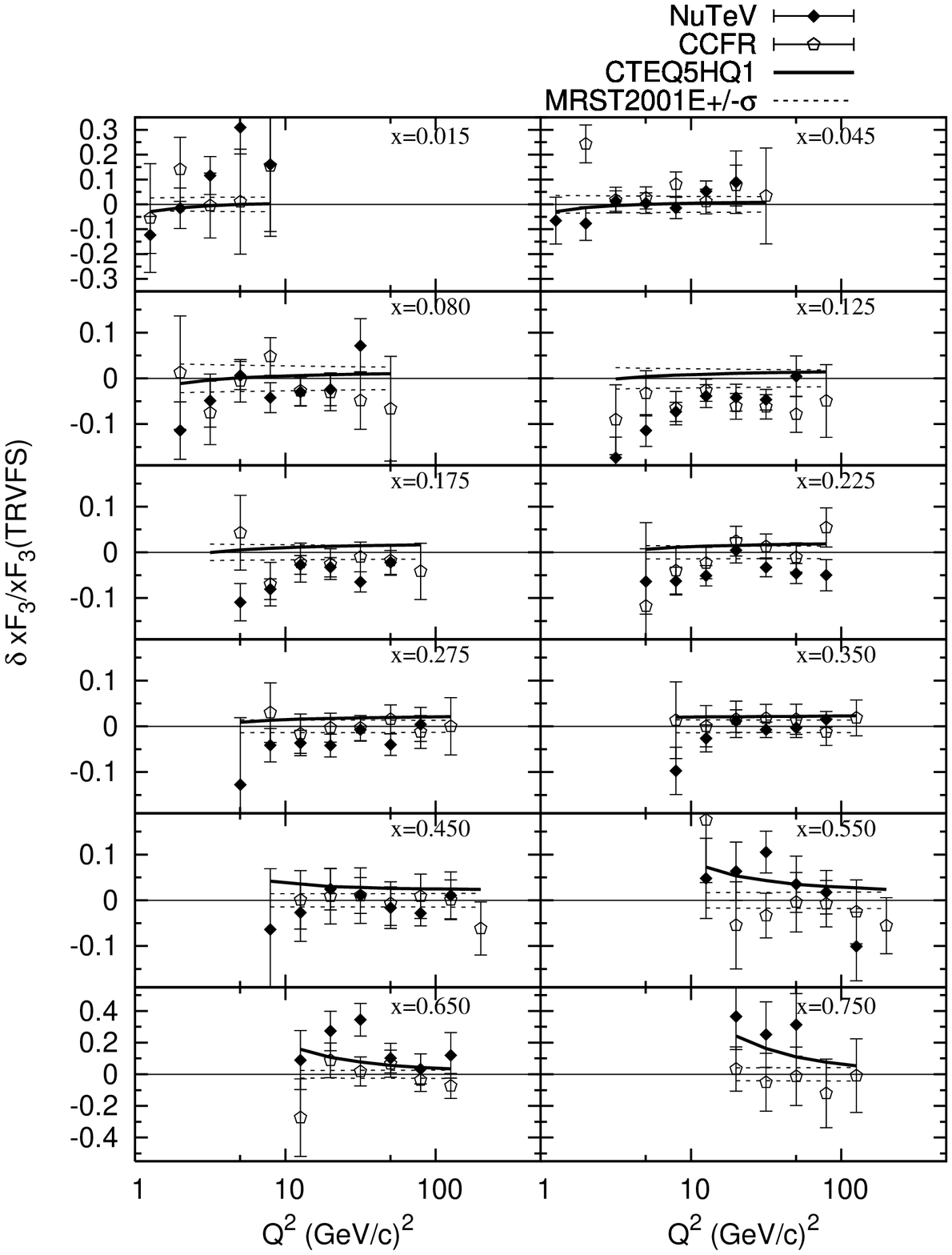}
  \caption{NuTeV and CCFR $F_2$(left) and $xF_3$(right) compared with TRVFS(MRST2001E) and ACOT(CTEQ5HQ1).}
 \label{f2th} 
\end{figure}

A comparison with TRVFS(MRST2001E)~\cite{trvfs,mrst2001e} 
and ACOT(CTEQ5)~\cite{acot,cteq5} for $F_2$ and $xF_3$ is shown
on Fig.\ref{f2th}. Both theoretical curves are corrected 
for nuclear target~\cite{nuRev,pdg} 
and target mass effects~\cite{gp_tm}.
NuTeV agrees with both theoretical calculations
for $0.06<x<0.5$. For $x<0.06$ both NuTeV and CCFR measure different
$Q^2$-dependence than the theoretical predictions. At high-$x$ both
theoretical predictions are systematically higher than the NuTeV
$F_2$ and $xF_3$.

The nuclear correction used to correct the theory curves is 
independent of $Q^2$ and based on a fit to charged-lepton data on 
nuclear targets. NuTeV perhaps indicates that neutrino scattering 
favors smaller nuclear effects at high-$x$ than are found in charged-lepton scattering. 
At small $x$, new theoretical calculations show that in the shadowing 
region the nuclear correction has $Q^2$ dependence~\cite{nuclear2,nuclear3}.
The standard nuclear correction obtained from a fit to charged lepton data implies a suppression of 10\% independent of $Q^2$ at $x=0.015$, while 
for $x=0.015$ reference~\cite{nuclear3} finds a suppression of 15\% at
$Q^2=1.25$GeV$^2$ and a suppression of 3.4\% at $Q^2=7.94$GeV$^2$.
This effect improves agreement with data at low-$x$.

\section{Conclusions}
In conclusion, NuTeV has measured $F_2$ and $xF_3$ structure functions.
This is the most precise measurement from neutrino scattering experiment to date. 
NuTeV result is in good agreement with previous $\nu$-Fe results over the intermediate $x$
region.  
% At high-$x$ ($x>0.4$) NuTeV result is systematically higher than CCFR result.
At high-$x$ NuTeV result is higher than the theoretical predictions. 
Perhaps, the nuclear correction is different for neutrino scattering.

\end{document}